\documentclass[11pt,aps, superscriptaddress,nofootinbib]{revtex4}
\usepackage{hyperref,amsmath,amsfonts,xcolor}

\begin{document}
\title{\bf \Large Axion Mass Bound in Very Special Relativity}
 
  \author{R. Bufalo} 
   \email{rodrigo.bufalo@dfi.ufla.br}
  \affiliation{Departamento de F\'isica, Universidade Federal de Lavras,
Caixa Postal 3037, 37200-000 Lavras, MG, Brazil}

   \author { S. Upadhyay }
   \email{ sudhakerupadhyay@gmail.com;  sudhaker@iitkgp.ac.in}
\affiliation {Centre for Theoretical Studies, Indian Institute of Technology
Kharagpur,  Kharagpur-721302, WB, India}

\begin{abstract} 
In this paper we propose a very special relativity (VSR)-inspired description of the axion electrodynamics.
This proposal is based upon the construction of a proper study of the SIM$(2)$--VSR gauge-symmetry.
It is shown that the VSR nonlocal effects give a health departure from the usual axion field theory.
The axionic classical dynamics is analysed in full detail, first by a discussion of its solution in the presence of an external magnetic field. Next, we compute photon-axion transition in VSR scenario by means of Primakoff interaction, showing the change of a linearly polarized light to a circular one. Afterwards, duality symmetry is discussed in the VSR framework.  
\end{abstract}

\date{\today}
\maketitle


\newpage

\tableofcontents

 \section{Introduction}

The axion is a hypothetical light and weakly interacting elementary particle postulated by Peccei--Quinn in 1977 as a solution to the strong CP problem in quantum chromodynamics (QCD) associated with a new $U(1)$ symmetry \cite{Peccei:1977hh,Weinberg:1977ma,Wilczek:1977pj}. 
Although it had been initially thought that the \emph{invisible} axion solves the strong CP problem without being amenable to verification by experiments,  
we have witnessed 40 years of intensive research on axion physics, based on either astrophysical observations or pure laboratory based experiments \cite{Turner:1989vc, Kuster:2008zz,Jaeckel:2010ni}.
So far, unfortunately, none was able to yield a positive signature for the axion or an axion-like particle.

Besides being originally proposed as a solution to the strong CP problem, axion-like particles plays an important part in explaining unanswered questions of cosmology \cite{Marsh:2015xka}. Moreover, due to its weakness of their interactions with a sufficiently small mass, axions go as one of the prominent candidates to account for the dark matter in the Universe \cite{Ringwald:2012hr,Ballesteros:2016euj}.

Notice, however, that non-trivial QCD vacuum effects (e.g., instantons) spoil the Peccei--Quinn symmetry explicitly and provide a small mass for the axion. Hence, the axion is viewed actually as a pseudo-Nambu-Goldstone boson \cite{Weinberg:1977ma,Wilczek:1977pj}, with a non-vanishing, but parameterically small mass. 
On the other hand, instead of considering the traditional Peccei-Quinn mechanism,
we will approach the axion dynamics from an alternative point of view,  where Lorentz violating effects are responsible to engender massive effects. 
In this sense, we shall focus in exploring features of  VSR  \cite{Cohen:2006ky,Cohen:2006ir} in this paper.

The cornerstone from the VSR proposal is that the laws of physics are not invariant under the whole Poincar\'e group but rather under subgroups of the
Poincar\'e group preserving the basic elements of special relativity, but at the same time enhancing the Lorentz algebra by modifying the dynamics of particles.
In particular, within this proposal, it is useful in the realization of VSR the use of representations of the full Lorentz
group but supplemented by a Lorentz-violating factor, such that the symmetry of the Lagrangian is then reduced to one of the VSR subgroups of the Lorentz group. 
These effects can then be encoded in the form Lorentz-violating terms in the Lagrangian that are necessarily nonlocal.

As an example, one can observe that a VSR-covariant Dirac equation has the form 
\begin{equation}
\left(i\gamma^\mu \tilde{\partial}_\mu -M \right)\Psi \left(x\right)=0,
\end{equation}
where the wiggle operator is defined such as $\tilde{\partial}_{\mu}=\partial_{\mu}+\frac{1}{2}\frac{m^{2}}{n.\partial}n_{\mu}$,
with the chosen preferred null direction $n_{\mu}=\left(1,0,0,1\right)$ so that it transforms multiplicatively
under a VSR transformation. So, by squaring the VSR-covariant Dirac equation we find
\begin{equation}
\left(\partial^\mu \partial_\mu +\mathcal{M}^2 \right)\Psi \left(x\right)=0, \quad \mathcal{M}^2=M^2+m^2 .
\end{equation}
We thus immediately realize that conservation laws and the usual relativistic dispersion relation are preserved; moreover, an interesting observable consequence of VSR that is to provide a novel mechanism for introducing neutrino masses without the need for new particles \cite{Cohen:2006ir}.
Moreover, the VSR parameter $m$ sets the scale for the VSR effects.
 
Let us now explain how axion dynamics can be defined in order to encompass Lorentz violating effects.
Due to its sensitive tests, photons are always good candidates as test particles in order to probe a physical system \cite{Adler:1971wn}.
In this sense, we can explore the fact that axions can be converted to photons and vise--versa in the presence of magnetic fields \cite{Sikivie:1983ip,Kaplan:1985dv,Srednicki:1985xd,Maiani:1986md,Gasperini:1987im} in order to detect modifications in the axion dynamics, more precisely to probe prominent VSR effects in the theory's dynamics in a significant and novel manner. 

As it concerns our interest, VSR-effects have been discussed in the context of electromagnetic theories:
Abelian and non-Abelian Maxwell theories \cite{Cheon:2009zx,Alfaro:2013uva}, Chern-Simons theory \cite{Vohanka:2014lsa,Vohanka:2015hha,Nayak:2015xba,Bufalo:2016lfq}, Born-Infeld electrodynamics \cite{Bufalo:2015gja} and higher-spin gauge fields \cite{Upadhyay:2015tkn,Upadhyay:2016hdj,Upadhyay:2017tkn}.
So, in this paper, we shall consider the conversion of photons into axions in the presence
of a background magnetic field, in the sense of Primakoff effect, where the VSR will play a part in the photon and axion sectors and will be responsible to engender massive nonlocal effects. It should be stressed that there are ongoing efforts in order to establish axion effects due to an electromagnetic probing \cite{Sikivie:2013laa,McAllister:2015zcz,Villalba-Chavez:2016hxw,Anastassopoulos:2017ftl}.

In this paper, we will examine the Axion electrodynamics in a VSR setting.
We start Sec.~\ref{sec2} by establishing the VSR-axion electromagnetic dynamics main aspects and reviewing the SIM$(2)$--VSR gauge invariance,
which allow us to determine the VSR-modified Abelian field-strength to be used in our analysis.
Moreover, we do first compute the solution for the axion field $\theta$ in the presence of an external magnetic field in terms of a plane wave solution.
In Sec.~\ref{sec3}, we compute explicitly the VSR photon-axion transition rate in a Primakoff framework, showing the change of a linearly polarized light to a circular one. Afterwards, in Sec.~\ref{sec4} duality symmetry for the VSR axion electrodynamics is established.  
In Sec.\ref{sec5} we summarize the results, and present our final remarks.

\section{VSR Axion mass}
\label{sec2}

We define the Lagrangian for the axion electrodynamics without source term in
VSR as given by 
\begin{align}
{\cal L}_{\rm axion}=-\frac{1}{4 }\tilde{F}_{\mu\nu}\tilde{F}^{\mu\nu}+\frac{\kappa}{4 }\theta\tilde{F}_{\mu\nu}\tilde{G}^{\mu\nu}+\frac{1}{2}\tilde{\partial}_{\mu}\theta\tilde{\partial}^{\mu}\theta, \label{eq1}
\end{align}
where $\kappa$ is the dimensionful parameter characterizing
the strength of the axion-photon coupling, $\theta(x)$ is a pseudo-scalar field known as the axion-like field, wiggle derivative is defined as before by $\tilde{\partial}_{\mu}=\partial_{\mu}+\frac{1}{2}\frac{m^{2}}{n\cdot\partial}n_{\mu}$, $\tilde{F}^{\mu\nu}$ and $\tilde{G}^{\mu\nu}=\frac{1}{2}\epsilon^{\mu\nu\rho\sigma}\tilde{F}_{\rho\sigma}$
are the field-strength the dual field-strength, respectively.
The axion electrodynamics in Lorentz invariant case admits a new internal (gauge) symmetry of the axion-electromagnetic field Lagrangian due to duality transformation \cite{VISINELLI:2013fia}, between the axion field and the gauge potential, which in turn leads to a conserved current. We discuss this point from a VSR perspective in later section \ref{sec4}.

Notice the absence of a potential for the axion field.
The $\theta\tilde{F}_{\mu\nu}\tilde{G}^{\mu\nu}$ term is responsible
to provide a solution the strong CP problem, known as Peccei-Quinn
solution. It is also known as the effective potential for the axion
field, and it is related to the axion mass $m_{a}^{2}=\left\langle \frac{\partial^{2}V_{\textrm{eff}}}{\partial\theta^{2}}\right\rangle $,
generated due to the spontaneous breaking of the $U\left(1\right)_{PQ}$
symmetry.
As discussed before, we replace this mechanism by VSR nonlocal point-of-view defined in \eqref{eq1}, which encompass Lorentz violating effects and are responsible to engender mass for the axion field in such a way that the axion mass has nothing to do with axion-photon coupling $\kappa$.

Besides, in order to write-down an expression for $F_{\mu\nu}$ we make use of
the usual definition of the raw field-strength $\left[D_{\mu},D_{\nu}\right]\phi=-iF_{\mu\nu}\phi$.
This is ensured by the construction of a gauge invariant quantity,
where the covariant derivative is given by \cite{Alfaro:2013uva}
\[
D_{\mu}\phi=\partial_{\mu}\phi-iA_{\mu}\phi+\frac{i}{2}m^{2}n_{\mu}\left(\frac{1}{\left(n\cdot\partial\right)^{2}}\left(n\cdot A\right)\right)\phi,
\]
which satisfy the transformation law $\delta\left(D_{\mu}\phi\right)=i\Lambda\left(D_{\mu}\phi\right)$,
where $\delta A_{\mu}=\partial_{\mu}\Lambda$.
On the other hand, the raw field-strength $F_{\mu\nu}$ does not coincide with the wiggle operator
\[
\tilde{F}_{\mu\nu}=\tilde{\partial}_{\mu}A_{\nu}-\tilde{\partial}_{\nu}A_{\mu}
\]
However, we can realize that the difference between the raw and wiggle field-strength must be gauge invariant as well. So that the wiggle in terms of the usual derivative can be written as \cite{Alfaro:2013uva}
\begin{align}
\tilde{F}_{\mu\nu}=\partial_{\mu}A_{\nu}+\frac{m^{2}}{2}n_{\mu}\left(\frac{1}{\left(n\cdot\partial\right)^{2}}\partial_{\nu}\left(n\cdot A\right)\right)-\mu\leftrightarrow\nu, \label{eq2}
\end{align}
which is gauge invariant and it will be used to describe massive gauge fields.

Lagrangian \eqref{eq1} will now be extensively explored in order to establish some features concerning axion physics, basically it describes how axions can be converted into photons, and vice versa. This basic process, known as Primakoff process, arising from the electromagnetic anomaly and expressed in the effective interaction
with coupling constant, underpins many constraints on axions.

First, we will determine solutions for the axion field equation in the presence of an external magnetic field, that can work as a source axion produced in laboratory due to the conversion of photons into axions, which might be seen as an inverse Primakoff process \cite{Gasperini:1987im}.
In the next section, Sec.~\ref{sec3}, we will discuss axion-photon interaction via direct Primakoff process, in which we observe the variation of the polarization state of a light wave interacting with the axion field in the presence of an external magnetic field \cite{Maiani:1986md}.

The sourceless dynamical field equations can be obtained from \eqref{eq1}, and for the electromagnetic and axion fields they read
\begin{align}
 & \tilde{\partial}_{\mu}\tilde{F}^{\mu\nu}-\kappa\tilde{G}^{\mu\nu}\tilde{\partial}_{\mu}\theta=0, \label{eq2a}\\
 & \left(\square+m^{2}\right)\theta=\frac{\kappa}{4 }\tilde{F}_{\mu\nu}\tilde{G}^{\mu\nu}, \label{eq2b}
\end{align}
where the differential identity $\tilde{\square}=\square+m^{2}$ and the Bianchi identity $\tilde{\partial}_{\mu}\tilde{G}^{\mu\nu}=0$
have been used.
Please notice the presence of massive excitations in \eqref{eq2b} that are engendered by VSR effects.
If we now make use of the definitions for the electric
and magnetic fields $E^{i}=\tilde{F}^{i0}$ and $B^{i}=\frac{1}{2}\epsilon^{ijk}\tilde{F}_{jk}$, respectively, we have
\begin{align}
 & \tilde{\nabla}\cdot \mathbf{E}-\kappa \mathbf{B}\cdot\tilde{\nabla}\theta=0, \label{eq3a}\\
 & \tilde{\nabla}\times \mathbf{B}-\tilde{\partial}_{0}\mathbf{E}+\kappa \mathbf{B}\tilde{\partial}_{0}\theta-\kappa\left(\mathbf{E}\times\tilde{\nabla}\theta\right)=0, \label{eq3b}\\
 & \left(\square+m^{2}\right)\theta=-\kappa\mathbf{B}\cdot\mathbf{E}, \label{eq3c}
\end{align}
In the VSR setting, $\tilde{\partial}_{\mu}\tilde{G}^{\mu\nu}=0$, the complementary electromagnetic field equations read
\begin{align}
 & \tilde{\nabla}\cdot\mathbf{B}=0,\\
 & \tilde{\nabla}\times\mathbf{E}+\tilde{\partial}_{0}\mathbf{B}=0.
\end{align}

In order to establish the framework of observing axions produced due to a electromagnetic wave we consider a strong uniform background magnetic field $\mathbf{B}_{0}$, orthogonal to the wave propagation. This can be achieved by means of
\begin{equation}
\tilde{\mathcal{F}}_{\mu\nu}= F_{\mu\nu}^{\rm ext}+ \tilde{F}_{\mu\nu}
\end{equation}
where $F_{\mu\nu}^{\rm ext}$ represents the external magnetic field.
In this context the field equations Eqs.~\eqref{eq3a}--\eqref{eq3c} are written as
\begin{align}
 & \left(\square+m^{2}\right)\mathbf{E}-\kappa\mathbf{B}_{0}\tilde{\partial}_{0}^{2}\theta=0, \label{eq4a} \\
 & \left(\square+m^{2}\right)\theta=-\kappa\mathbf{B}_{0}\cdot\mathbf{E}, \label{eq4b}
\end{align}
It is important to emphasize that here, the axion mass $m$ is entirely due to VSR effects, and has nothing to do with axion-photon coupling $\kappa$. Furthermore, in this setting, Eqs.~\eqref{eq4a} and \eqref{eq4b}, both photon and axion fields have the same mass, displaying screened profiles.

A simple setup to determine the solution for the axion field is to take it propagating
along the $\hat{x}$ direction, $\phi\left(x,t\right)$. Moreover,
we can decompose the electric field $\mathbf{E}$ into components
perpendicular $E_{\bot}$ and parallel $E_{\Vert}$
to the external field $\mathbf{B}_{0}$, respectively. Within this framework,
we get the following coupled field equations
\begin{align}
 & \left(\square+m^{2}\right)E_{\bot}=0,\label{eq5a}\\
 & \left(\square+m^{2}\right)E_{\Vert}+\kappa B_{0}\tilde{\partial}_{0}^{2}\theta=0 \label{eq5b}\\
 & \left(\square+m^{2}\right)\theta=-\kappa E_{\Vert}B_{0}. \label{eq5c}
\end{align}
Notice that the perpendicular component $E_{\bot}$ do not
couple to the axion field.

By simplicity, we can also consider that the background magnetic field
$\mathbf{B}_{0}$ is limited to a region $0\leq x\leq L$,
while is vanishing outside this region. In this case, we easily see
that we can represent the axion field as free plane waves in the noninteracting
regions \cite{Gasperini:1987im}.

Now in the interacting region we can make use of the plnce wave decomposition
\begin{equation}
E_{\Vert}=E_{0}e^{-i\left(\omega t-kx\right)},\quad\theta=\theta_{0}e^{-i\left(\omega t-kx\right)},
\end{equation}
We then get
\begin{align}
& \left(-\omega^{2}+k^{2}+m^{2}\right)E_{0}-\kappa\tilde{\omega}^{2}B_{0}\theta_{0}=0 \label{eq6a}\\
 & \left(-\omega^{2}+k^{2}+m^{2}\right)\theta_{0}+\kappa E_{0}B_{0}=0,  \label{eq6b}
\end{align}
where we have defined $\tilde{\omega}^{2}=\omega^{2}-\frac{\omega m^{2}}{n\cdot k}+\frac{m^{4}}{4\left(n\cdot k\right)^{2}}$.
A solution to the axion field $\theta $ in this case read
\begin{equation}
\theta_{0}\left(\omega,k\right)=\kappa\frac{E_{0}B_{0}}{\omega^{2}-\omega_{m}^{2}}, \label{eq6}
\end{equation}
where $\omega_{m}=\pm\sqrt{k_m^{2}+m^{2}}$. It is important to notice
that this solution exists provided $\omega$ and $\omega_{m}$ satisfy
the condition
\begin{align}
 & \left(-\omega^{2}+\omega_{m}^{2}\right)\left(\omega^{2}-\omega_{m}^{2}\right)-\kappa^2B_{0}^{2}\tilde{\omega}^{2}=0 \label{eq7}
\end{align}
In particular, if we realize that $n\cdot k=\omega$ we can write
\begin{align}
\omega^{6}-a\omega^{4}+b\omega^{2}+c=0 \label{eq7b}
\end{align}
where we have identified $a=2\omega_{m}^{2}-\kappa^{2}B_{0}^{2}$,
$b=\omega_{m}^{4}-\kappa^{2}B_{0}^{2}m^{2}$
and $c=\frac{\kappa^{2}}{4 }B_{0}^{2}m^{4}$.

From the three dispersion relation solutions for \eqref{eq7b}, two of them have an
imaginary part, showing a damped behavior of the plane wave in the given
region in both solutions. The real and complex dispersion relations read
\begin{align}
\omega^{2}  &=\frac{a}{3}-\frac{2^{\frac{1}{3}}}{3 \Xi^{\frac{1}{3}}} \left(3b-a^{2}\right)+ \frac{\Xi^{\frac{1}{3}}}{32^{\frac{1}{3}}} \label{eq8},  \\
\omega^{2} _{\pm } &= \frac{a}{3}- \frac{1\pm i\sqrt{3}}{6 2^{\frac{1}{3}}} \Xi^{\frac{1}{3}} + \frac{1\pm i\sqrt{3} }{3 2^{\frac{2}{3}} \Xi^{\frac{1}{3}}}  \left(3b-a^{2}\right)
\end{align}
where $\Xi=2a^{3}-9ab-27c+3\sqrt{3}\sqrt{27c^{2}+18abc-4a^{3}c+4b^{3}-a^{2}b^{2}}$.

So general solutions for the axion field, satisfying \eqref{eq5c}, in the free and interacting regions with the background magnetic field are
\begin{equation}
\begin{cases}
\theta_I\left(t,x\right)=c_{1}e^{-i\left(\omega_{m}t+kx\right)}, \\
\theta_{II}\left(t,x\right)=c_{2}e^{-i\left(\omega_{m}t+kx\right)}+c_{3}e^{-i\left(\omega_{m}t-kx\right)}+\kappa\frac{E_{0}B_{0}}{\omega^{2}-\omega_{m}^{2}}e^{-i\left(\omega t-kx\right)}, \\
\theta_{III}\left(t,x\right)=c_{4}e^{-i\left(\omega_{m}t-kx\right)} .
\end{cases}\label{eq9a}
\end{equation}
All the amplitudes in this solution can be uniquely obtained
by imposing the continuity conditions at $x=0$ and $x=L$, \footnote{Notice that in these expressions we have made use of the notation in terms of the wave numbers $k$ and $k_m$ instead of frequencies $\omega$ and $\omega_m$. }
\begin{align}
\begin{cases}
c_{1} & =\frac{1}{2}\frac{\kappa E_{0}B_{0}}{k_{m}\left(k+k_{m}\right)}\left(e^{i\left(k+k_{m}\right)L}-1\right),\\
c_{2} & =\frac{1}{2}\frac{\kappa E_{0}B_{0}}{k_{m}\left(k+k_{m}\right)}e^{i\left(k+k_{m}\right)L},\\
c_{3} & =-\frac{1}{2}\frac{\kappa E_{0}B_{0}}{k_{m}\left(k-k_{m}\right)},\\
c_{4} & =\frac{1}{2}\frac{\kappa E_{0}B_{0}}{k_{m}\left(k-k_{m}\right)}\left(e^{i\left(k-k_{m}\right)L}-1\right) .
\end{cases}
\end{align}
The axion field solutions \eqref{eq9a} represent waves outgoing from the interaction region.

A possible use of the solutions \eqref{eq9a} is to get an estimate of the axion flux density attainable from an artificial source \cite{Gasperini:1987im}.
This can be achieved by solving the axion and ``electric field'' coupled equations iteratively, Eqs. \eqref{eq6a} and \eqref{eq6b}, starting from the decoupled equations and considering the first-order contribution in $\kappa$ onto the axion field. This results into
\begin{equation}
J  \simeq \frac{E_{e}\kappa^{2}B_{0}^{2}\omega_m}{m^4}, \label{eq10}
\end{equation}
where $E_{e}=E_{0}^{2}/2$  is the irradiance of a linearly polarized electromagnetic wave (Poynting vector).

In turn, Eq.\eqref{eq10} can be applied to estimate possible production by electromagnetic fields of general axion-like particles, and to collect general information on axion parameters, e.g. their masses and coupling constants. However, it should be notice that within VSR framework both photon and axion masses have the same origin and are strictly the same, i.e. due to Lorentz violating effects in VSR, see Eqs.\eqref{eq5a}--\eqref{eq5c}.

In particular, it should be emphasized that bounds on photon mass $m_\gamma \leq 1.8 \times 10^{-14}~{\rm eV}$ \cite{Bonetti:2016cpo} are much more stronger than those in axion mass $m_a \leq 1 \times 10^{-2}~{\rm eV}$ \cite{Raffelt2008}. Since,  both the masses have the same signature, the stringent bound on photon mass can also be imposed onto the axion's as well.

\section{VSR photon-axion transition}
\label{sec3}

We turn our attention to the analysis of photon production due to an axion source.
In the presence of a magnetic field, the Primakoff interaction between
axions and photons allows for the vacuum to become birefringent and
dichroic \cite{Maiani:1986md,Deffayet:2001pc}. These effects cause the polarization plane of linearly polarized light to be rotated as it propagates.

In order to investigate the phenomenon, we proceed to compute the
photon-axion conversion rate in the VSR scenario. For this matter,
we return to the equations of motion \eqref{eq2a} and \eqref{eq2b}, but written now in terms of the vector potential $\mathbf{A}$ instead of the electric
field $\mathbf{E}$. In our analysis we consider the radiation gauge, $A_0 =0$ and $\nabla \cdot \mathbf{A} =0$. Thus, keeping only linear terms in $\mathbf{A}$ and $\theta$, the classical field equations are written as
\begin{align}
\left(\square+m^{2}\right)\mathbf{A}+\kappa \mathbf{B}\tilde{\partial}_{0}\theta & =0 \label{eq10a}\\
\left(\square+m^{2}\right)\theta-\kappa \mathbf{B}\cdot\tilde{\partial}_{0}\mathbf{A} & =0 \label{eq10b}
\end{align}

Moreover in the small perturbations $\omega\approx k$ regime (i.e.
a WKB limit where we assume that the amplitude varies slowly), we
find the linearized system of equations
\begin{align}
\left(-2\left(\omega+i\partial_{x}\right)+m^{2}\right)A_{\perp} & =0 \label{eq11a}\\
\left(-2\left(\omega+i\partial_{x}\right)+m^{2}\right)A_{\Vert}-i\kappa\tilde{\omega}B_{0}\theta & =0 \label{eq11b}\\
\left(-2\left(\omega+i\partial_{x}\right)+m^{2}\right)\theta+i\kappa\tilde{\omega}B_{0}A_{\Vert} & =0 \label{eq11c}
\end{align}
where we have once again decomposed the potential $\mathbf{A}$ into components
perpendicular $A_{\bot}$ and parallel $A_{\Vert}$ to the external field.
If we introduce a vector
\begin{equation}
\Psi=\left(\begin{array}{c}
\theta_{0}\left(x\right)\\
A_{\Vert}\left(x\right)\\
A_{\bot}\left(x\right)
\end{array}\right)e^{-i\omega x} \label{eq12}
\end{equation}
 and identify $\tilde{\omega}=\omega-\frac{m^{2}}{2\omega}$, we can rewrite the above equations, Eqs.\eqref{eq11a}--\eqref{eq11c}, in a more suitable Schrödinger-like form
\begin{equation}
i\frac{d}{dx}\Psi=\left(\begin{array}{ccc}
\frac{m^{2}}{2\omega} & -i\frac{\kappa\tilde{\omega}B_{0}}{2\omega} & 0\\
i\frac{\kappa\tilde{\omega}B_{0}}{2\omega} & \frac{m^{2}}{2\omega} & 0\\
0 & 0 & \frac{m^{2}}{2\omega}
\end{array}\right)\Psi=\mathbf{M}\Psi \label{eq13}
\end{equation}
where we have defined $\mathbf{M}$ as the mixing matrix. 

In order to highlight the VSR effects, let us consider the photon-axion
conversion by considering that the axion can only convert in the parallel
component $A_{\Vert}$. Hence, 
\begin{equation}
i\frac{d}{dx}\left(\begin{array}{c}
\theta_{0}\left(x\right)\\
A_{\Vert}\left(x\right)
\end{array}\right)=\left(\begin{array}{cc}
\frac{m^{2}}{2\omega} & -i\frac{\kappa\tilde{\omega}B_{0}}{2\omega}\\
i\frac{\kappa\tilde{\omega}B_{0}}{2\omega} & \frac{m^{2}}{2\omega}
\end{array}\right)\left(\begin{array}{c}
\theta_{0}\left(x\right)\\
A_{\Vert}\left(x\right)
\end{array}\right)=\left(\begin{array}{cc}
\Delta_{a} & -i\Delta_{m}\\
i\Delta_{m} & \Delta_{\parallel} 
\end{array}\right)\Phi \label{eq14}
\end{equation}
Usually the diagonal term involving the vector potential $A_{\Vert}$
is related to its effective mass due to the Euler-Heisenberg effective
Lagrangian, plasma effect (since, in general, the photon does not
propagate in vacuum) and Cotton-Mouton effect, i.e. the birefringence
of gases and liquids in presence of a magnetic field, so that $\Delta_{\parallel}=\Delta_{VSR}+\Delta_{EH}+\Delta_{\textrm{plasma}}+\Delta_{CM}$   \cite{Deffayet:2001pc},
that is it receives further contribution than the VSR one $\Delta_{VSR}=\frac{m^{2}}{2\omega}$.
However, we can see that a photon effective mass is naturally encompassed
in the VSR framework, as well as the axion. On the other hand, this might be seen as its
bare mass, being corrected by further effects as mentioned.

Now, to compute the photon-axion conversion probability we must first
diagonalize the above mixing matrix, whose eigenvalues read
\begin{equation}
\chi_{\pm}=\frac{\left(\Delta_{a}+\Delta_{\parallel}\right)\pm\sqrt{\left(\Delta_{a}-\Delta_{\parallel}\right)^{2}+4\Delta_{m}^{2}}}{2} \label{eq15}
\end{equation}
However, in the bare case, i.e. taking into account solely VSR effects,
we have that 
\begin{equation}
\Delta_{a}=\Delta_{\parallel}=\frac{m^{2}}{2\omega}\equiv\Delta.
\end{equation}
This give us the following simple relation
\begin{equation}
\chi_{\pm}=\Delta\pm\Delta_{m} \label{eq16}
\end{equation}
in which we can assume that the above matrix is diagonalized through
an orthornormal transformation $\tilde{\Phi}=\mathbf{O}\Phi$, or
even $\mathbf{O}^{\dagger}\mathbf{M}_{\parallel}\mathbf{O}=\mathbf{M}_{D}$,
\begin{equation}
\mathbf{O}=\left(\begin{array}{cc}
\cos\varphi & \sin\varphi\\
-\sin\varphi & \cos\varphi
\end{array}\right)
\end{equation}
where $\varphi$ is the mixing angle. We thus obtain the following
solution for the axion $\theta_{0}$ and photon $\tilde{A}_{\Vert}$
fields
\begin{align}
\theta_{0}\left(x\right) & =\left(\cos^{2}\varphi e^{-i\chi_{+}x}+\sin^{2}\varphi e^{-i\chi_{-}x}\right)\theta_{0}\left(0\right)+\cos\varphi\sin\varphi\left(e^{-i\chi_{+}x}-e^{-i\chi_{-}x}\right)A_{\Vert}\left(0\right) \label{eq17a}\\
A_{\Vert}\left(x\right) & =\cos\varphi\sin\varphi\left(e^{-i\chi_{+}x}-e^{-i\chi_{-}x}\right)\theta_{0}\left(0\right)+\left(\sin^{2}\varphi e^{-i\chi_{+}x}+\cos^{2}\varphi e^{-i\chi_{-}x}\right)A_{\Vert}\left(0\right) \label{eq17b}
\end{align}

With the solutions in hands, we can easily compute the probability
of oscillation of a photon after make a distance $x$ starting from the
initial state, in which we consider the initial state as $\theta_{0}\left(0\right)=0$
and $A_{\Vert}\left(0\right)=1$. Hence, the photon-axion
conversion probability can be evaluated as
\begin{equation}
P\left(\gamma\rightarrow a\right)=\left|\left\langle A_{\Vert}\left(0\right)|\theta_{0}\left(x\right)\right\rangle \right|^{2}=\sin^{2}\left(2\varphi\right)\sin^{2}\left(\frac{\left(\chi_{+}-\chi_{-}\right)x}{2}\right) \label{eq18}
\end{equation}
We can characterize the transition by introducing the oscillation
length $\ell_{\textrm{osc}}=2\pi/\Delta_{\textrm{osc}}$, where the
oscillation wavenumber reads
\begin{equation}
\Delta_{\textrm{osc}}=\chi_{+}-\chi_{-}
\end{equation}
Some remarks are in place. Now, in general, if we had $\Delta_{a}\neq\Delta_{\parallel}$, we would have
\begin{equation}
\Delta_{\textrm{osc}}=\sqrt{\left(\Delta_{a}-\Delta_{\parallel}\right)^{2}+4\Delta_{m}^{2}}=\frac{2\Delta_{m}}{\sin\left(2\varphi\right)} \label{eq19}
\end{equation}
Notice that a complete transition between a photon and an axion is
only possible when the mixing is maximal, i.e. when $\varphi=\pi/4$.
However, in our case, notice that due to VSR effects we have the equality
$\Delta_{a}=\Delta_{\parallel}$, which implies
\begin{equation}
\sin\left(2\varphi\right)=\frac{2\Delta_{m}}{\sqrt{\left(\Delta_{a}-\Delta_{\parallel}\right)^{2}+4\Delta_{m}^{2}}}=1
\end{equation}
This shows that in the VSR framework we naturally have the strong
mixing regime: $\Delta_{\textrm{osc}}=2\Delta_{m}$. Hence, the transition
rate \eqref{eq18} is written in its final form
\begin{equation}
P\left(\gamma\rightarrow a\right)=\sin^{2}\left(\Delta_{m}x\right)\sim\left(\Delta_{m}x\right)^{2} \label{eq20}
\end{equation}

Since only one of the photon components can mix with the axion, in our case $A_{\parallel}$, so the photon-axion conversion can affect polarization of the photon. This can be analyzed by means of the Stokes parameters \cite{McMaster:1961}
\begin{align*}
I\left(x\right) & =A_{\Vert}\left(x\right)A_{\Vert}^{*}\left(x\right)+A_{\perp}\left(x\right)A_{\perp}^{*}\left(x\right)\\
Q\left(x\right) & =A_{\Vert}\left(x\right)A_{\Vert}^{*}\left(x\right)-A_{\perp}\left(x\right)A_{\perp}^{*}\left(x\right)\\
U\left(x\right) & =A_{\Vert}\left(x\right)A_{\perp}^{*}\left(x\right)+A_{\perp}\left(x\right)A_{\Vert}^{*}\left(x\right)\\
V\left(x\right) & =i\left(A_{\Vert}\left(x\right)A_{\perp}^{*}\left(x\right)-A_{\perp}\left(x\right)A_{\Vert}^{*}\left(x\right)\right)
\end{align*}
The degree of polarization can be readily defined in terms of such
parameters, the circular polarization reads
\begin{equation}
\Pi_{C}=\frac{\left|V\left(x\right)\right|}{I\left(x\right)}
\end{equation}
while the linear polarization is
\begin{equation}
\Pi_{L}=\frac{\sqrt{Q^{2}\left(x\right)+U^{2}\left(x\right)}}{I\left(x\right)}
\end{equation}
In the VSR case, we have that the photon circular polarization is
\begin{equation}
\Pi_{C}=\left|\frac{U_{0}\left(\Delta_{m}/2\right)}{I_{0}-\frac{1}{4}\left(I_{0}+Q_{0}\right)\left(1-\cos\left(2\Delta_{m}x\right)\right)}\left[\frac{\sin\left(\chi_{+}x\right)}{\chi_{+}}-\frac{\sin\left(\chi_{-}x\right)}{\chi_{-}}\right]\right| \label{eq21}
\end{equation}
where $I_{0},Q_{0},U_{0}$ represent the initial Stokes parameters,
we took $V_{0}=0$ in the above relation.

We then explicitly see that in the VSR framework the change of linearly polarized light to be rotated in the Eq.~\eqref{eq21} is due to the Primakoff interaction.

\section{Duality symmetry and conserved current  for VSR Axion}
\label{sec4}

In order to conclude our discussion, we shall now present an analysis of the duality symmetry of the axion electrodynamics \cite{VISINELLI:2013fia} but now in the VSR framework.
It is well known that the sourceless dynamical field equation for the electromagnetic field and their complementary field equations without axion field is invariant under $SO(2)$ rotation by an angle $\zeta$, 
\begin{equation}
\left(\begin{array}{c}
\tilde F^{\prime \mu\nu} \\
\tilde G^{\prime \mu\nu}
\end{array}\right)=\left(\begin{array}{cc}
\cos \zeta & \sin\zeta\\
-\sin\zeta & \cos \zeta
\end{array}\right)\left(\begin{array}{c}
\tilde F^{ \mu\nu} \\
\tilde G^{ \mu\nu}
\end{array}\right).
\end{equation}
Now, to analyze the duality symmetry in the VSR modified axion electrodynamics Eq.~\eqref{eq1}, we apply the $SO(2)$ transformation with $\zeta=\pi/2$ in Eq.~\eqref{eq2a}
\begin{equation}
\tilde{\partial}_{\mu}\tilde{G}^{\mu\nu}-\kappa\tilde{F}^{\mu\nu}\tilde{\partial}_{\mu}\theta=0.
\end{equation}
This leads to the new following set of field equations
\begin{align}
&\tilde{\nabla}\cdot \textbf{B}+\kappa \textbf{E}\cdot \tilde{\nabla}\theta =0,\nonumber\\
&\tilde{\nabla}\times \textbf{E}+\kappa (\textbf{B}\times   \tilde{\nabla}\theta)+\tilde{\partial}_0 \textbf{B}+\kappa \textbf{E}\tilde{\partial}_0 \theta =0.
\end{align}

We can obtain the gauge field equations of VSR axion electrodynamics theory
by defining the electric and magnetic fields in terms of a new gauge potential $\hat{A}_\mu$ as
\begin{align}
\textbf{B}+\kappa\theta \textbf{E}&\equiv \hat{\textbf{B}}=\tilde{\nabla}\times \hat{\textbf{A}},\nonumber\\
\textbf{E}-\kappa\theta \textbf{B}&\equiv\hat{\textbf{E}}=-\tilde{\partial}_0\hat{\textbf{A}}-\tilde{\nabla}\hat{A}0.
\end{align}
Notice the use of the wiggle derivatives in the above definition.
In order to study the conserved current of axion electrodynamics in VSR, we consider that the vector field has the following configuration
 \begin{equation}
 \textbf{A}=A_x \hat i+B_x y\hat k,
 \end{equation}
where  $A_x$ depends on time only, while $B_x$ is a constant and equal to the magnitude
of the magnetic field, $\tilde {\nabla} \times  \textbf{{A}}=B_x\hat i$.
In this scenario, the action describing the axion
electrodynamics for $\theta$ and the field $A_x$ in VSR is \footnote{Notice, however, that the mass term here is due to VSR effects, i.e. $\tilde{\partial}_\mu\phi\tilde{\partial}^\mu\phi = \partial_\mu\phi\partial^\mu\phi -m^2 \phi ^2$.}
\begin{align}
S=\frac{1}{2}\int d^4x\left[\partial_\mu\theta\partial^\mu\theta -m^2\theta^2+
\partial_\mu A_x\partial^\mu A_x -m^2 A_x^2 +\kappa B_x(\theta\dot{A}_x-A_x \dot 
{\theta})
\right].\label{act}
\end{align}
The equations of motion for  $A_x$ and $\theta$ are given, respectively, by
\begin{align}
(\square +m^2)A_x+kB_x {\partial_0}\theta &=0,\nonumber\\
(\square +m^2)\theta-\kappa B_x {\partial_0}A_x &=0.
\end{align}
Here, we observe that the mass terms appear naturally due to VSR effects, without the need of a potential term.
The action \eqref{act}  is invariant under the following gauge symmetry:
\begin{align}
\delta \theta &=A_x\eta,\nonumber\\
\delta A_x &=-\theta\eta, \label{sym} 
\end{align}
where $\eta$ is an infinitesimal  (dimensionless) constant parameter.
It is important to emphasize that in the usual framework the transformations \eqref{sym} are a symmetry of the action \eqref{act} only when one of the two conditions are satisfied: either i) photon have a bare mass, equal to the axion mass, or ii) photon and axion fields are massless \cite{VISINELLI:2013fia}.
Notice that the first condition is naturally satisfied in the VSR framework, showing hence that VSR axion electrodynamics has duality symmetry by construction.

At last, utilizing the Noether' theorem, we are able to calculate conserved 
charge and current. These are
\begin{align}
J^0&=\theta {\partial_0}A_x -A_x {\partial_0}\theta +\frac{\kappa}{2}B_x(A_x^2+\theta^2),\nonumber\\
J^i&=\theta( {\partial^i} A_x)-( {\partial^i}\theta) A_x.
\end{align}
From the above expressions, the conservation of current $\partial_\mu J^\mu =0$ is evident.

 \section{Conclusion}
 \label{sec5}
 In this paper, we have studied a VSR inspired modification of the axion electrodynamics.
The analysis consisted in first formulation a SIM$(2)$--VSR axion electrodynamics, 
with the expectation that the nonlocal (Lorentz violating) effects would contribute in a novel way showing a distinct departure from the usual theory. Due to the results obtained, a natural extension of the present analysis would be a study concerning QCD, more precisely the strong CP problem in the VSR setting, where the Lorentz violating effects might play an interesting part in the Peccei-Quinn mechanism.

We started with a brief review on the VSR formalism for the Abelian gauge sector so that we have a proper formulation of the VSR axion electrodynamics. 
In order to extract physical features of the model, we have chosen to exploit Primakoff interaction, i.e. the photon-axion transition. First, we have considered the inverse Primakoff process, the production of axions due to a photons source. In this case, we have fully established the axion field solution in the presence of an external magnetic field.

Next, we have considered the production of photons due to axions source, more precisely we computed the photon-axion conversion probability. In particular, we have shown that in the VSR framework we naturally have the strong mixing regime, i.e. the maximum production of photons due to axions. Besides, we have computed the photon circular polarization by means of Stokes parameters, showing in the Primakoff process in a VSR framework the change of linearly polarized light to a circular one.

At last, we have discussed the duality symmetry in the VSR setting. It is remarkable to notice that due the fact that both photon and axion acquire the same mass $m$ due to VSR effects, showing thus that VSR axion electrodynamics is by construction invariant by duality symmetry.


\subsection*{Acknowledgments}

R.B. gratefully acknowledges CNPq for partial support, Project No. 304241/2016-4.

\end{document}